\documentclass[twocolumn,showpacs,pre,aps]{revtex4}

\usepackage{dcolumn}
\usepackage{bm}

\begin{document}

\title{From Shock Waves to Brownian Motion and 1/f\,-\,Noise in Gas}

\author{Yuriy E. Kuzovlev}
\email{kuzovlev@kinetic.ac.donetsk.ua} \affiliation{Donetsk Institute
for Physics and Technology (DonPTI NASU), 83114 Donetsk, Ukraine}


\begin{abstract}
A formally exact relation is derived which connects thermodynamically
non-equilibrium evolution of gas density distribution after its
arbitrary strong spatially non-uniform perturbation and evolution of
many-particle correlations between path of some marked particle and
its surroundings in equilibrium gas. This relation directly confirms
significance of the many-particle correlations even under the
Boltzmann-Grad limit and thus validates the earlier suggested
revision of kinetics.
\end{abstract}


\maketitle

\section{Introduction}
The ``law of large numbers'', first discovered by Bernoulli
approximately 300 years ago \cite{jb}, up to now is world outlook
paradigm of physical applications of probabilities. Meanwhile, at the
end of his work \cite{jb} Bernoulli himself emphasized that if
equivalent observations were perpetually continued (and at last
probability turned to certainty) then one would conclude that all
things obey exact rules or even fate. In the field of probabilistic
interpretation of quantum mechanics such conclusion means
``nonlocality'' not less impressive than the famous EPR paradox.
Naturally, one may ask who is counting outcomes of our observations
(e.g. castings a die) and makes so that their relative frequencies
have definite limits equal to speculative ``probabilities'' (e.g. all
equal to $\,1/6\,$) ?

The standard answer is that those godlike role belongs to
``independency'' of the outcomes, which claims that probability of
their series is product of elemental probabilities. But this is
tautology, not the answer, because ``elemental probability'' loses
meaning in absence of the limit. A good explanation of this logical
fact can be found in \cite{sad}. Hence, presence (or absence) of the
limit is nothing but postulate concerning probabilistic measure in a
space of observation sequences, and no more.

Unfortunately, in statistical physics built on microscopic dynamics
the only a priori probabilities at our disposal are probabilities of
its initial conditions. In such the theory, as Krylov emphasized
almost 60 years ago \cite{kr}, we should distinguish between actual
dynamical (cause-and-effect) dependencies or correlations and
statistical ones. Moreover, sometimes they controvert one to another.
The determinism of dynamics can produce infinitely long-living actual
correlations and thereby arrange definite limits for relative
frequencies being perceived as fast decay of statistical correlations
and ``independency'' of observations (similarly, ideality of a die is
incarnate actual correlation ensuring ``independency'' of its
castings). Many such examples are known from mixing (time-reversible
chaotic) nonlinear dynamics with finite number of degrees of freedom.

And, vice versa, deficiency of actual dynamical dependencies may
result in arbitrary long statistical correlations. For example, if
earlier collisions of a particle (or charge carrier) with gas of
similar particles (or with phonon gas) have no influence on later
ones, then such system is indifferent to what rate (relative
frequency) of collisions does happen (therefore it would be vain to
wait till ``probability turns to certainty''). The result is
scaleless fluctuations, possessing 1/f\,-type spectrum, in collision
rate of a test particle and thus in its diffusivity and mobility
(1/f\,-noise) \cite{bk1,bk2,i1} (for more details and examples see
also \cite{bk3,i2,p1}).

As it was shown in \cite{i1} (see also \cite{i2,p1,p2}), in such case
one has no rights to introduce a priori ``probabilities of
collisions'', all the more, postulate ``molecular chaos''. Instead,
we should honestly investigate dynamical evolution of an initial
statistical ensemble, taking in mind Krylov's conjecture that under
mixing dynamics, generally, relative frequencies of some phenomenon
along particular phase trajectories in no way are connected with some
its a priory probability \cite{kr}.

In case of gas, if wishing to investigate Brownian motion of its
particles, we have to solve the BBGKY equations starting from
spatially non-uniform statistical ensemble with certainly known
(non-random) position of at least one (test) particle. An exact
solution is wittingly impossible, but we can presume the gas not too
dense and resort to approximative shortened description of collisions
in terms of Boltzmannian ``collision integrals''. However, analysis
of this approach in \cite{i1} revealed that the Boltzmann's
``molecular chaos'' hypothesis is incompatible with spatial
non-uniformity of the ensemble. As the consequence, in place of the
single Boltzmann equation the theory births infinite chain of kinetic
equations. They operate with usual one-particle distribution function
plus infinite set of specific many-particle distribution functions
which represent clusters of (real or virtual) collisions or close
encounters of particles and together describe fluctuations in the
rate of collisions \cite{i1,i2,p1}.

In view of innovatory character of these specific functions, it seems
principal to make sure that they have adequate prototypes beyond
approximations. Below we present a proof based on exact relations
between such functions and a relaxation law for arbitrary strong
perturbations of gas density.

\section{Start statistical ensemble}
First, we will consider statistical ensemble of $\,N\,$ particles
which are contained in a volume $\,\Omega\,$ and possess short-range
repulsion of one from another, under two successive formal limits:\,
firstly the ``thermodynamical'' one, $\,N\rightarrow\infty\,$,
$\,\Omega\rightarrow\infty\,$, $\,\nu_{\,0} =N/\Omega
\rightarrow\,$const\,, and secondly the ``Boltzmann-Grad limit'',
$\,\nu_{\,0}\rightarrow\infty\,$, $\,\sigma\rightarrow 0\,$,
$\,\lambda =(\sigma\nu_{\,0})^{-1}\rightarrow\,$const\,
($\,\lambda\,$ is mean free path and $\,\sigma\,$ properly defined
cross-section of inter-particle collisions).

Let initially, at time moment $\,t=0\,$, our system is described by
the $\,N$-particle distribution function
\begin{eqnarray}
\rho^{(in)}(\Gamma)\,=\,C^{(in)}\,\rho^{(eq)}(\Gamma)\,
\exp{[\,-\sum_{j\,=1}^N
V(\bm{r}_j)/T\,]}\,\,\,\,, \nonumber \\
\rho^{(eq)}(\Gamma)\,=\,C^{(eq)}\,\exp{\left [\,-
\,H(\Gamma\,)/T\,\right
]}\,\,\,\,\,\,\,\,\,\,\,\,\,\,\,\,\,\,\label{fin}
\end{eqnarray}
Here $\,\Gamma =\{\Gamma_1,...\,,\Gamma_N\}\,$ with
$\,\Gamma_j=\{\bm{r}_j\,,\bm{p}_j\}\,$ being variables of $\,j$-th
particle,\, $\,H\,$ denotes full $\,N$-particle Hamiltonian,\,
$\,\rho^{(eq)}\,$ is corresponding canonical equilibrium distribution
,\, $\,V(\bm{r})\,$ is a bounded function, and constants
$\,C^{(eq)}\,$ and $\,C^{(in)}\,$ ensure the normalization
\begin{eqnarray}
\int \rho^{(in)}(\Gamma)\,\prod_{j\,=1}^Nd\Gamma_j\,=\,\int
\rho^{(eq)}(\Gamma)\,\prod_{j\,=1}^Nd\Gamma_j\,=\,1 \label{norm}
\end{eqnarray}

We will be interested in consequent evolution of the system, that is
in distribution function
\begin{eqnarray}
\rho(t,\Gamma)\,=\,\exp{(\,tL\,)}\,\rho^{(in)}(\Gamma)\,\,\,,
\,\,\,\,\,\,\,\,\,\,\,\,\,\,\,\,\,\,\,\,\,\,\,\,\,\label{ft}\\
L\,=\,\sum_{j\,=1}^N\, [\,(\,\partial H/\partial
\,\bm{r}_j)\,\partial/\partial \bm{p}_j\,-\,(\partial H/\partial
\bm{p}_j)\,\partial/\partial
 \,\bm{r}_j\,]\,\,,\nonumber
\end{eqnarray}
where $\,L\,$ is the Liouville operator corresponding to Hamiltonian
$\,H\,$. Clearly, $\,\rho^{(in)}\,$ and $\,\rho(t)\,$ represent
statistical ensemble which is non-equilibrium because of non-uniform
perturbation of particles density
\begin{eqnarray}
\nu(t,\bm{r})\,=\,\int\,\rho(t,\Gamma)\,\sum_{j\,=1}^N
\delta(\bm{r}-\bm{r}_j)\,\prod_{j\,=1}^Nd\Gamma_j \label{den}
\end{eqnarray}
Alternatively, one can think that $\,\rho(t)\,$ comes from
equilibrium in presence of an external potential $\,V(\bm{r})\,$
which however is abruptly switched off at $\,t=0\,$.

It is convenient to characterize density perturbations by function
\begin{eqnarray}
\phi(\bm{r})\,\equiv\,\exp{[-V(\bm{r})/T\,]}\,-1\,
\,\nonumber
\end{eqnarray}
Then, introducing functional
\begin{eqnarray}
\mathcal{F}\{\phi\}\,\equiv\,
\int\,\rho^{(eq)}(\Gamma)\,\prod_{j\,=1}^N
[\,1+\phi(\bm{r}_j)\,]\,\,d\Gamma_j\,\,\,\,,\label{norm1}
\end{eqnarray}
one can easy verify the equalities
\begin{eqnarray}
C^{(in)}\,=\,1/\mathcal{F}\{\phi\}\,\,\,\,,\,\,\,\,
\,\,\,\,\,\,\,\,\,\,\,\,\,\,\,\,\,\,\,\,\,\,\,\,\,\label{cin}\\
\nu(0,\bm{r})\,=\,[\,1+\phi(\bm{r})\,]\, \frac
{\delta\ln\mathcal{F}\{\phi\}}{\delta\,\phi(\bm{r})} \label{iden}
\end{eqnarray}

In order to perform the thermodynamical limit, we will assume that as
a whole the perturbation is finite:
\begin{eqnarray}
\int |\phi(\bm{r})|\,d\bm{r}\,<\,\infty \label{n0}
\end{eqnarray}
Then the limit of the functional (\ref{norm1}) looks as
\begin{eqnarray}
\mathcal{F}\{\phi\}\,=\,1+\sum_{k\,=1}^{\infty}\frac
{\nu_{\,0}^{\,k}}{k!}\,\int F_k(...)\,\prod_{j\,=1}^k\phi(\bm{r}_j)
\,d\bm{r}_j\,=\nonumber \\
=\,\,\exp\,\sum_{k\,=1}^{\infty}\frac {\nu_{\,0}^{\,k}}{k!}\,\int
\Upsilon_k(...)\,\prod_{j\,=1}^k\phi(\bm{r}_j)
\,d\bm{r}_j\,\,\,\,,\,\,\,\,\,\,\,\,\,\label{func}
\end{eqnarray}
where $\,F_k(...)=F_k(\bm{r}_1,...\,,\bm{r}_k)\,$ are usual
non-normalized (or, in other words, ``normalized to volume'')
$\,k$-particle spatial distribution functions (or, in other words,
``correlation functions'') of strictly equilibrium gas. In
particular, $\,F_1(\bm{r}_1)=1\,$.

Functions $\,\Upsilon_k(...)=\Upsilon_k(\bm{r}_1,...,\bm{r}_k)\,$
formally relate to $\,F_k\,$ like cumulants, or semi-invariants, of a
random field relate to its statistical moments:\,
$\,\Upsilon_2(\bm{r}_1,\bm{r}_2)=$ $\,F_2(\bm{r}_1,\bm{r}_2)-
F_1(\bm{r}_1)F_1(\bm{r}_2)= F_2(\bm{r}_1,\bm{r}_2)-1\,$\,, and so on.

In fact, of course, all $\,F_k\,$ and $\,\Upsilon_k\,$ obey
translation invariance, i.e. depend on inter-particle distances
$\,\bm{r}_i-\bm{r}_j\,$ only. At that, any of the cumulants
$\,\Upsilon_k\rightarrow 0\,\,$ when some of its arguments
$\,\bm{r}_i-\bm{r}_j\,$ goes to infinity as measured in units of
characteristic radius $\,\,r_0\,\,$ of particles interaction (taking
in mind 3-D gas, one can define $\,\,r_0\,\,$ e.g. by means of
$\,\sigma =\pi r_0^{\,2}\,$). Therefore, if the function
$\,\phi(\bm{r})\,$ is smooth enough, on scales of order of $\,r_0\,$,
then
\begin{eqnarray}
\int \Upsilon_k\prod_{j\,=1}^k\phi(\bm{r}_j)
\,d\bm{r}_j\,\rightarrow\,\gamma_k\,r_0^{\,3(k-1)}\,
\int\phi^{\,k}(\bm{r}_1)\,d\bm{r}_1\,\,\label{ups0}
\end{eqnarray}
at $\,k>1\,$ with $\,\gamma_k\,$ being ``virial coefficients''. Under
the Boltzmann-Grad limit, $\,r_0\rightarrow 0\,$ and the ``gas
parameter'', defined e.g. by $\,g=(4\pi r_0^{\,3}/3)\,\nu_{\,0}\,$\,,
also tends to zero, $\,g\rightarrow 0\,$. Consequently, gas becomes
ideal from the point of view of thermodynamics:
\begin{eqnarray}
\mathcal{F}\{\phi\}\,\rightarrow\,\exp\left [\,\nu_{\,0}
\int\phi(\bm{r})\,d\bm{r} \right ]\,\,\,\,,\,\label{fl0}\\
\frac {\nu(0,\bm{r})}{\nu_{\,0}}\,\rightarrow\,1+\phi(\bm{r})\,=
\,\exp{[\,-V(\bm{r})/T\,]}\,\,\label{bgl0}
\end{eqnarray}

\section{Relaxation, time symmetry and
equilibrium-non-equilibrium relations} However, from the point of
view of kinetics our gas remains quite non-ideal. Let us consider
relations which substitute (\ref{iden}) and (\ref{func}) at $\,t>0\,$
and connect, on one hand, non-equilibrium process of relaxation of
relative gas density perturbation, $\,\nu(t,\bm{r})/\nu_{\,0}\,-1\,$,
and, on the other hand, Brownian motion of particles in unperturbed
equilibrium gas.

Taking into account that the Liouville operator $\,L\,$ commutes with
any equilibrium distribution, $\,L\,\rho^{(eq)} f=\rho^{(eq)}L f\,$
(where $\,f\,$ is ``arbitrary function'' of $\,\Gamma_j\,$), and
combining formulas (\ref{fin}), (\ref{ft}) and (\ref{den}), we can
transform the latter to
\begin{eqnarray}
\nu(t,\bm{r})\,=\,\frac
{1}{\mathcal{F}\{\phi\}}\,\int\prod_{j\,=1}^Nd\Gamma_j\,
\,\,\rho^{(eq)}(\Gamma)\,\,\times\label{den1} \\
\times\,\,\sum_{j\,=1}^N
\delta(\bm{r}-\bm{r}_j)\,\exp{(\,tL)}\,\prod_{j\,=1}^N
\,[\,1+\phi(\bm{r}_j)\,]\,\nonumber
\end{eqnarray}
Let us make the change of integration variables here:
$\,\bm{p}_j\Rightarrow -\,\bm{p}_j\,$. Since the Hamiltonian is even
function (quadratic form) of the momentums while $\,L\,$ is odd with
respect to them, the only result of this operation is change of sign
of time argument before $\,L\,$:
\begin{eqnarray}
\nu(t,\bm{r})\,=\,\frac
{1}{\mathcal{F}\{\phi\}}\,\int\prod_{j\,=1}^Nd\Gamma_j\,\,
\,\rho^{(eq)}(\Gamma)\,\times\,\,\,\,\,\,\,\label{den2} \\
\times\,\,\sum_{j\,=1}^N
\delta(\bm{r}-\bm{r}_j)\,\exp{(\,-\,tL)}\,\prod_{j\,=1}^N\,
[\,1+\phi(\bm{r}_j)\,]\, \,\,\,\,\,\,\,\,\nonumber
\end{eqnarray}
which clearly demonstrates time symmetry of gas evolution:
$\,\nu(-t,\bm{r})=\nu(t,\bm{r})\,$.

Next, recall that operator $\,\exp{(\,-tL)}\,$ has twofold sense: if
acting onto distribution functions it describes their time-reversed
evolution, but if acting onto system's variables (or any function of
them) it describes their evolution in real direct time. In
particular,
\begin{eqnarray}
\exp{(\,-\,tL)}\,\,\phi(\bm{r}_j)\,=\,
\phi(\bm{r}_j(t,\Gamma))\,\,\,\,\,,\,\label{var}
\end{eqnarray}
where expression $\,\bm{r}_j(t,\Gamma)\,$ means (vector of)
coordinates of $\,j$-th particle at time $\,t\,$ considered as a
function of initial state $\,\Gamma\,$ of the whole system taken at
$\,t=0\,$ (i.e. the set $\,\bm{r}_j(t,\Gamma)\,$ is solution to
system's Hamiltonian equations starting from
$\,\bm{r}_j(0,\Gamma)=\bm{r}_j\,$,
$\,\bm{p}_j(0,\Gamma)=\bm{p}_j\,$)\,.

Hence, obviously, we can rewrite (\ref{den2}) in the form
\begin{eqnarray}
\frac {\nu(t,\bm{r})}{\nu_{\,0}}\,=\,\int\frac
{\rho^{(eq)}(t,\Gamma\,|\,\bm{r})}{\mathcal{F}\{\phi\}}\,
\prod_{j\,=1}^N\,
[\,1+\phi(\bm{r}_j)\,]\,d\bm{\Gamma}_j\,\,\label{den3}
\end{eqnarray}
where $\,\rho^{(eq)}(t,\Gamma\,|\,\bm{r})\,$ is conditional
distribution function of system's state $\,\Gamma\,$ at current time
$\,t\,$ corresponding to condition that initially (at time zero) one
of particles was positioned exactly at a given point $\,\bm{r}\,$.
Formally,
\begin{eqnarray}
\rho^{(eq)}(t,\Gamma\,|\,\bm{r})\,=\,
\exp{(\,tL)}\,\rho^{(eq)}(0,\Gamma\,|\,\bm{r})\,\,\,\,,
\,\,\,\,\,\,\nonumber\\
\rho^{(eq)}(0,\Gamma\,|\,\bm{r})\,=\Omega\,\rho^{(eq)}(\Gamma\,)\,\frac
1N \sum_{j\,=1}^N \delta(\bm{r}-\bm{r}_j)\,\,\,, \label{roin}
\end{eqnarray}
where factor $\,\Omega/N\,$ ensures normalization to unit:
$\,\int\rho^{(eq)}(t,\Gamma\,|\,\bm{r})\,d\Gamma =1\,$, and we
already keep in mind the coming thermodynamical limit.

To complete the limit, notice that all of $\,N\,$ terms of
(\ref{roin}) give equal contributions to integral in (\ref{den3}). In
analogy with transition from (\ref{norm1}) to (\ref{func}), the
result is
\begin{eqnarray}
\frac {\nu(t,\bm{r})}{\nu_{\,0}}\,=\,\frac
{1}{\mathcal{F}\{\phi\}}\,\int\,d\bm{r}^{\prime}\,\,
[\,1+\phi(\bm{r}^{\prime})\,]\,\{\,W_1(t,\bm{r}^{\prime}|\bm{r})+
\,\,\,\nonumber \\
+\sum_{k\,=1}^{\infty}\frac {\nu_{\,0}^{\,k}}{k!}\int
F_{k+1}(t,\bm{r}^{\prime}|\bm{r}_1,...\,,\bm{r}_k|\bm{r})
\prod_{j\,=1}^k\phi(\bm{r}_j)d\bm{r}_j\,\}\,\,\,\, \label{rfunc}
\end{eqnarray}

Here $\,W_1(t,\bm{r}^{\prime}|\bm{r})\,$ means probability density of
finding at point $\,\bm{r}^{\prime}\,$ at time $\,t\,$ the same test
particle what started from the point $\,\bm{r}\,$. This function is
normalized to unit:
\begin{equation}
\begin{array}{c}
\int W_1(t,\bm{r}^{\prime}|\bm{r})\,d\bm{r}^{\prime}\,=1\,\,\,
\label{wn}
\end{array}
\end{equation}
In other words, this is just the distribution of Brownian
displacement of test particle which was under investigation in
\cite{p1} (and much earlier in \cite{i1,i2}).

Functions
$\,F_{k+1}(t,\bm{r}^{\prime}|\bm{r}_1,...\,,\bm{r}_k|\bm{r})\,$
describe joint probabilities of finding the test particle (started
from $\,\bm{r}\,$) at point $\,\bm{r}^{\prime}\,$ and simultaneously
some $\,k\,$ other particles at points $\,\bm{r}_j\,$. These
functions are symmetric with respect to $\,\bm{r}_j\,$
($\,j=1...\,k\,$) and definitely normalized with respect to
$\,\bm{r}^{\prime}\,$ as well as $\,\bm{r}\,$:
\begin{eqnarray}
\int F_{k+1}(t,\bm{r}^{\prime}|\bm{r}_1,...\,,\bm{r}_k|\bm{r})\,
d\bm{r}^{\prime}\,=\,F_{k}(t,\bm{r}_1,...\,,\bm{r}_k|\bm{r})
\,\,\,\,\, \label{n}\\
\int F_{k+1}(t,\bm{r}^{\prime}|\bm{r}_1,...\,,\bm{r}_k|\bm{r})\,
d\bm{r}\,=\,F_{k+1}(\bm{r}_1,...\,,\bm{r}_k,\bm{r}^{\prime})
\,\,\,\,\, \label{nn}
\end{eqnarray}
On right-hand side in (\ref{n}) we see non-normalized (i.e.
``normalized to whole volume'') distribution functions (correlation
functions) characterizing probability of location of $\,k\,$
particles at points $\,\bm{r}_j\,$ under condition that one more
particle had started from $\,\bm{r}\,$. Functions
$\,F_{k+1}(\bm{r}_1,...\,,\bm{r}_k,\bm{r}^{\prime})\,$ on right-hand
side of (\ref{nn}) are usual static $\,(k+1)$-particle distribution
functions mentioned in previous Section.

Of course, after the limit all these functions acquire translation
invariance:
$\,W_1(t,\bm{r}^{\prime}|\bm{r})=W_1(t,\bm{r}^{\prime}-\bm{r})\,$,
etc. With this notion it becomes rather obvious that identities
(\ref{nn}) guarantee ``conservation of particles'': $\,\int
\nu(t,\bm{r})\,d\bm{r}\,=\,\int \nu(0,\bm{r})\,d\bm{r}$.

Initial conditions for these functions also are clear:
\begin{equation}
\begin{array}{c}
W_1(0,\bm{r}^{\prime}|\bm{r})\,=\,
\delta(\bm{r}^{\prime}-\bm{r})\,\,\,,\label{ic}\\
F_{k+1}(0,\bm{r}^{\prime}|\bm{r}_1,...\,,\bm{r}_k|\bm{r})\,
=\,\delta(\bm{r}^{\,\prime}-\bm{r})\,
F_{k+1}(\bm{r}_1,...\,,\bm{r}_k,\bm{r})\,\,\,,\\
F_{k}(0,\bm{r}_1,...\,,\bm{r}_k|\bm{r})\,
=\,F_{k+1}(\bm{r}_1,...\,,\bm{r}_k,\bm{r})\,\,\,
\end{array}
\end{equation}
Correspondingly, at $\,t=0\,$ (\ref{rfunc}) reduces to (\ref{iden})
with (\ref{func}).

In principle, (\ref{rfunc}) can be qualified as example of
``generalized Onsager relations'' or ``generalized
fluctuation-dissipation relations'' which follow from time
reversibility of microscopic dynamics and establish connections
between dissipative nonlinearity and statistics of transport
processes, in particular, statistics of equilibrium noise (see e.g.
\cite{fds,p} and references therein).

\section{Dynamical multi-particle correlations and their cumulants}
The distribution functions what appear in (\ref{rfunc}) mix up i)
average values, ii) static correlations between \,{\it positions}\,
of particles due to their potential interaction, and besides iii)
specific dynamical correlations between current relative disposition
of mutually colliding (or merely encountering) particles and their
previous \,{\it displacements}\,. Next, we would like to separate
those contributions and especially consider the third of them.

In order to extract joint dynamical correlations in a pure form, we
should introduce cumulants, or semi-invariants, similar to cumulant
functions $\,\Upsilon_k(\bm{r}_1,...,\bm{r}_k)\,$ in (\ref{func}).
With this purpose, first, notice that expression inside braces in
(\ref{rfunc}) has the same formal structure as expression $\,\langle
\,x\,\exp{(\phi\,y)}\,\rangle\,$, where $\,x\,$ and $\,y\,$ are two
random quantities, $\,\phi\,$ is dummy parameter (test parameter),
and angle brackets denote averaging procedure. Second, apply the well
known general formula (see e.g. \cite{mal})
\begin{widetext}
\begin{eqnarray}
\langle \,x\,\exp{(\phi\,y)}\,\rangle\,=\,\langle
\,\exp{(\phi\,y)}\,\rangle\left[\,\langle x\rangle+\phi\,\langle
x,y\rangle + \frac {\phi^2}{2!}\,\langle x,y,y\rangle +\frac
{\phi^3}{3!}\,\langle x,y,y,y\rangle+...\,\right
]\,\,\,\,\,\,\,\,\,\,\,\,\,\,\,\,\,\,\,\,\,\,\,\,\,
\,\,\,\,\,\,\,\,\,\,\,\,\,\,\,\,\,\,\label{c}
\end{eqnarray}
with $\,\langle\,x,y,...\,,y\,\rangle\,$ being joint cumulants in
Malakhov's notations \cite{mal}.

In our context, similar transformation as performed in (\ref{rfunc})
yields
\begin{eqnarray}
W_1(t,\bm{r}^{\,\prime}|\bm{r})+ \sum_{k\,=1}^{\infty}\frac
{\nu_{\,0}^{\,k}}{k!}\int
F_{k+1}(t,\bm{r}^{\,\prime}|\bm{r}_1,...\,,\bm{r}_k|\bm{r})
\prod_{j\,=1}^k\phi(\bm{r}_j)\,d\bm{r}_j\,=
\,\,\,\,\,\,\,\,\,\,\,\,\,\,\,\,\,\,\,\,\,\,\,
\,\,\,\,\,\,\,\,\,\,\,\,\,\,\,\,\,\,\,\,\,\,\,\,\,\,\,
\,\,\,\,\,\,\,\,\,\,\,\,\,\,\,\,\,\,\,\,\,\,\,
\,\,\,\,\,\,\,\,\,\,\,\,\,\,\,\nonumber \\
=\,\left[\,1+\sum_{k\,=1}^{\infty}\frac {\nu_{\,0}^{\,k}}{k!}\int
F_{k}(t,\bm{r}_1,...\,,\bm{r}_k|\bm{r})
\prod_{j\,=1}^k\phi(\bm{r}_j)\,d\bm{r}_j \right]\,\times
\,\,\,\,\,\,\,\,\,\,\,\,\,\,\,\,\,\,\,\,\,\,\,\,\,\,\,\,\,\,
\,\,\,\,\,\,\,\,\,\,\,\,\,\,\,\,\,\,\,\,\,\,\,\,\,\,\,\,\,\,
\,\,\,\,\,\,\,\,\,\,\,\,\,\,\,\,\,\,\,\,\,\,\,\,\,\,
\,\,\,\,\,\,\,\,\,\,\,\,\,\,\label{wdef}\\
\,\,\,\,\,\,\,\,\,\,\,\,\,\,
\times\,\left[\,W_1(t,\bm{r}^{\,\prime}|\bm{r})+
\sum_{k\,=1}^{\infty}\frac {\nu_{\,0}^{\,k}}{k!}\int
W_{k+1}(t,\bm{r}^{\,\prime}|\bm{r}_1,...\,,\bm{r}_k|\bm{r})
\prod_{j\,=1}^k\phi(\bm{r}_j)\,d\bm{r}_j\,\right]\,\equiv
\,\,\mathcal{F}\{t,\phi\,|\bm{r}\}\,\times\,
\mathcal{W}\{t,\bm{r}^{\,\prime}|\phi\,|\bm{r}\}\,\,\,\,\nonumber
\end{eqnarray}
Here new functions
$\,W_{k+1}(t,\bm{r}^{\prime}\,|\bm{r}_1,...\,,\bm{r}_k|\bm{r})\,$, in
place of
$\,F_{k+1}(t,\bm{r}^{\prime}\,|\bm{r}_1,...\,,\bm{r}_k|\bm{r})\,$,
are just the cumulant functions (analogous to
$\,\langle\,x,y,...\,,y\,\rangle\,$ in (\ref{c})) what correspond to
refined dynamical (two-time) inter-particle correlations.

Further, notice that because of the normalization relations
(\ref{wn}) and (\ref{n}) result of integration of the above
expression over $\,\bm{r}^{\,\prime}\,$ coincides with functional
$\,\mathcal{F}\{t,\phi\,|\bm{r}\}\,$ defined by the upper square
brackets in (\ref{wdef}). Therefore, firstly,
\begin{eqnarray}
\int\,\mathcal{W}\{t,\bm{r}^{\,\prime}|\phi\,|\bm{r}\}
\,\,d\bm{r}^{\prime}\,=\,1\,\,\,\,,\,\,\,\,\,\text{that is}
\,\,\,\,\,\,\int\,
W_{k+1}(t,\bm{r}^{\,\prime}|\bm{r}_1,...\,,\bm{r}_k|\bm{r})
\,\,d\bm{r}^{\prime}\,=\,0\,\,\,\,,\,\label{wp}
\end{eqnarray}
with functional
$\,\mathcal{W}\{t,\bm{r}^{\,\prime}|\phi\,|\bm{r}\}\,$ defined by the
lower square brackets in (\ref{wdef}). Secondly, formula
(\ref{rfunc}) takes the form
\begin{eqnarray}
\frac {\nu(t,\bm{r})}{\nu_{\,0}}\, =\,\frac
{\mathcal{F}\{t,\phi\,|\bm{r}\}}{\mathcal{F}\{\phi\}}\,\left
[\,1\,+\int
\phi(\bm{r}^{\,\prime})\,\mathcal{W}\{t,\bm{r}^{\,\prime}|\phi\,|\bm{r}\}
\,\,d\bm{r}^{\prime}\,\right ]\,\,\label{rfunc1}
\end{eqnarray}
Notice also that, because of the normalization relation (\ref{nn}),
integration of (\ref{wdef}) over $\,\bm{r}\,$ results in identity
\begin{eqnarray}
\int\mathcal{F}\{t,\phi\,|\bm{r}\}\,\,
\mathcal{W}\{t,\bm{r}^{\,\prime}|\phi\,|\bm{r}\}\,\,d\bm{r}\,=\,\frac
{1}{\nu_{\,0}}\,\,\frac {\delta \mathcal{F}\{\phi\}}{\delta\,
\phi(\bm{r}^{\,\prime})}\,
\,\,\,\,\,\,\,\,\,\,\,\,\,\,\,\,\,\,\,\,\,\,\,\,\,\label{id}
\end{eqnarray}
Besides, combining (\ref{wdef}) with initial conditions (\ref{ic}) we
obtain one more property of cumulants $\,W_{k+1}\,$:
\begin{eqnarray}
\mathcal{W}\{0,\bm{r}^{\,\prime}|\phi\,|\bm{r}\} \,=\,\delta
(\bm{r}^{\,\prime}-\bm{r})\,\,\,\,,\,\,\,\,\,\text{that is}
\,\,\,\,\,\,\,\,\,
W_{k+1}(0,\bm{r}^{\,\prime}|\bm{r}_1,...\,,\bm{r}_k|\bm{r})
\,=\,0\,\,\,\,,\,\label{icw}
\end{eqnarray}
thus confirming that indeed they represent correlations of very
dynamical nature.

Functions $\,F_{k}(t,\bm{r}_1,...\,,\bm{r}_k|\bm{r})\,$ also can be
replaced by corresponding cumulants if we write
\begin{eqnarray}
\mathcal{F}\{t,\phi\,|\bm{r}\}\,=\,\,\exp\,\sum_{k\,=1}^{\infty}\frac
{\nu_{\,0}^{\,k}}{k!}\int
\Upsilon_k(t,\bm{r}_1,...\,,\bm{r}_k|\bm{r})\prod_{j\,=1}^k\phi(\bm{r}_j)
\,d\bm{r}_j\,\,\,\,,\,\,\,\,\,\label{ff}
\end{eqnarray}
\end{widetext}
with $\Upsilon_k(t,\bm{r}_1,...\,,\bm{r}_k|\bm{r})\,$ being direct
analogue of above introduced static cumulant functions
$\,\Upsilon_k(\bm{r}_1,...,\bm{r}_k)\,$.

Let us discuss and compare two just now introduced sorts of
cumulants. Last of them,
$\,\Upsilon_k(t,\bm{r}_1,...\,,\bm{r}_k|\bm{r})\,$, at $\,k>1\,$
describe conditional joint correlations between $\,k\,$ particles
under condition that some one more $\,(k+1)$-th (test) particle
earlier, at $\,t=0\,$, was localized at point $\,\bm{r}\,$.
Certainly, such poor condition can not change short-range character
of the correlations, and the latter must decay at same inter-particle
distances $\,|\bm{r}_i-\bm{r}_j|\gtrsim \,r_0\,$ as unconditioned
correlations. Then, quite as in (\ref{ups0}),
\begin{eqnarray}
\int \Upsilon_k(t,\bm{r}_1,...\,,\bm{r}_k|\bm{r})
\prod_{j\,=1}^k\phi(\bm{r}_j)
\,d\bm{r}_j\,\approx\,\,\,\,\,\,\,\,\,\,\,\,\,\,\,\,\,
\,\,\,\,\,\,\,\,\,\,\,\,\,\,\,\,\,\,\,\nonumber\\
\approx\, r_0^{\,3(k-1)}\int \gamma_k(t,\bm{r}_1|\bm{r})\,
\phi^{\,k}(\bm{r}_1)\,d\bm{r}_1\,\,\,\,\,\,\,\,\label{ups}
\end{eqnarray}
All the more, such condition can not influence upon mean gas density.
Therefore it is natural to state, in agreement with (\ref{ic}), that
difference between the conditional one-particle distribution
function,
$\,\Upsilon_1(t,\bm{r}_1|\bm{r})=F_1(t,\bm{r}_1|\bm{r})\,$\,(expressing
conditional gas density in units of $\,\nu_{\,0}\,$), and the
unconditioned one, $\,\Upsilon_1(\bm{r}_1)=F_1(\bm{r}_1)=1\,$\,, at
$\,t>0\,$ is not significantly greater than at $\,t=0\,$. We can
write this in the form
\begin{eqnarray}
\left|\,\int [\,\Upsilon_1(t,\bm{r}_1|\bm{r})-\Upsilon_1(\bm{r}_1)\,]
\,\phi(\bm{r}_1)\,d\bm{r}_1\,\right|\,
\lesssim\,\,\,\,\,\,\,\,\,\,\,\,\,\,\,\,\,\,\,\,\label{hyp}\\
\,\,\,\,\lesssim \,\,\left|\,\int
[\,F_2(\bm{r}_1,\bm{r})-1\,]\,\phi(\bm{r}_1)\,d\bm{r}_1\,\right|\,
\sim\,r_0^3\,\overline{\phi} \,\,\,\,,\,\nonumber
\end{eqnarray}
where $\,\overline{\phi}\,$ is characteristic magnitude of
$\,\phi(\bm{r})\,$.

Formal substantiation of these statements in principle could be
obtained from the Bogolyubov-Born-Green-Kirkwood-Yvon (BBGKY)
equations
\begin{eqnarray}
\frac {\partial F^{(eq)}_n}{\partial
t}\,=\,L^{(n)}F^{(eq)}_n+\nu_{\,0}\sum_{j\,=1}^n\int
L_{j\,n+1}F^{(eq)}_{n+1}\,d\Gamma_{n+1}\,\,\,\,\,\,\label{bfn}
\end{eqnarray}
for the full phase space $\,n$-particle distributions
$\,F^{(eq)}_{n}=\,$
$\,=F^{(eq)}_{n}(0,\Gamma_1|\Gamma_2,...\,,\Gamma_{n-1}|\bm{r})\,$
connected to our spatial distributions by means of
\begin{eqnarray}
W_{1}(t,\bm{r}_1|\bm{r})\,=\, \int\,F^{(eq)}_{1}(t,\Gamma_1|\bm{r})
\,d\bm{p}_{\,1}\,\,\,\,,
\,\,\,\,\,\,\,\,\nonumber\\
F_{k+1}(t,\bm{r}_1|\bm{r}_2,...\,,\bm{r}_{k+1}|\bm{r})\,=\,\,\,
\,\,\,\,\,\,\,\,\,\,\,\,\,\,\,\,\,\,\,\,\,\label{rv}\\
=\,\int F^{(eq)}_{k+1}(t,\Gamma_1|\Gamma_2,...\,,\Gamma_{k+1}|\bm{r})
\,\,\prod_{j\,=1}^{k+1}d\bm{p}_j\,\,\, \nonumber
\end{eqnarray}
along with (\ref{n}), and subjected to initial conditions
\begin{eqnarray}
F^{(eq)}_{n}(0,\Gamma_1|\Gamma_2,...\,,\Gamma_{n-1}|\bm{r})\,=\,
\,\,\,\,\,\,\,\,\,\,\,\,\,\,\,\,\,\,\nonumber\\
=\,\delta(\bm{r}_1-\bm{r})\,F_{n}(\bm{r}_1,...\,,\bm{r}_n)\,
\prod_{j\,=1}^n G_0(\bm{p}_j)\,\,\,\,,\label{bic}
\end{eqnarray}
with ``number one'' assigned to the test particle, $\,L^{(n)}\,$
standing for $\,n$-particle Liouville operator,
\begin{eqnarray}
L_{\,ij}\,=\, \nabla U(\bm{r}_i-\bm{r}_j)\cdot\left(\partial
/\partial \bm{p}_{\,i}\,-\,\partial /\partial
p_{\,j}\,\right)\,\,\,\,,\nonumber
\end{eqnarray}
$\,U(\bm{R})\,$ being interaction potential, and $\,G_0(\bm{p})\,$
equilibrium Maxwellian momentum distribution. Clearly, this is a task
about Brownian motion of the test particle in equilibrium gas. But,
unfortunately, such BBGKY equations still remain almost
uninvestigated beyond roughening like the Boltzmannian kinetics or
its improvement suggested in \cite{i1,i2} (otherwise essays like the
present one would be unneeded).

Other conditional cumulants,
$\,W_{k+1}(t,\bm{r}^{\prime}\,|\bm{r}_1,...\,,\bm{r}_k|\bm{r})\,$,
seem much more interesting objects because describe joint
$\,(k+1)$-order correlations between $\,k\,$ ``ordinary'' particles
and besides the test particle itself. The latter is peculiar since it
is the only particle we know wherefrom it came to its current
position. Therefore, in fact $\,W_{k+1}\,$ represent joint
correlations between its preceding displacement
$\,\bm{r}^{\prime}-\bm{r}\,$ (``Brownian path'') and its current
surroundings. Moreover, such extra correlations what do not reduce to
mere gas density fluctuations already taken into account by
$\,\Upsilon_k\,$. Thus we must conclude that cumulants
$\,W_{k+1}(t,\bm{r}^{\prime}\,|\bm{r}_1,...\,,\bm{r}_k|\bm{r})\,$
represent fluctuations in the line of collisions of the test
particle, and in this sense, in contrary to $\,\Upsilon_k\,$, they
are specific dynamical correlations. Correspondingly, coordinates
$\,\bm{r}_j\,$ belong to those neighbors of the test particle which
currently are or recently were or soon will be involved into
collisions with it (or even those which would be in collision if some
other neighbor did not interfere with it).

As joint cumulants in general,
$\,W_{k+1}(t,\bm{r}^{\prime}\,|\bm{r}_1,...\,,\bm{r}_k|\bm{r})\,$
vanish if one (or more) of its constituents become statistically
independent on others. The independency, in its turn, takes place
when any of the relative distances
$\,|\bm{r}_j-\bm{r}^{\,\prime}\,|\,$ exceeds some characteristic
spatial scale $\,l_0\,$. However, the above reasoning prompts that
this scale is not the interaction radius $\,r_0\,$, figuring in
(\ref{ups0}) and (\ref{ups}), but ``radius of collision'' which is
sooner the mean free path $\,\lambda\,$ than $\,r_0\,$.

Indeed, clearly, information about a collision propagates to as long
distance apart it as $\,\lambda\,$, if not greater. Nevertheless,
$\,\lambda\,$ would be overestimate of $\,l_0\,$, since that is
propagation along the direction of relative momentum
$\,\bm{p}_j-\bm{p}^{\,\prime}\,$ (and velocity) only. At two
perpendicular dimensions the collision occupies merely a region
$\,\approx\,\sigma\sim r_0^2\,$. Hence, total ``volume of collision''
(and volume covered by related correlations) is \,$\,\approx\,
\sigma\lambda\,=\,1/\nu_{\,0}\,$. Correspondingly, it is reasonable
to estimate mean (averaged over all directions) ``radius of
collision'' (and radius of related correlations) $\,l_0\,$ as
$\,l_0\,=\,1/\nu_{\,0}^{1/3}\,$. In other words, $\,l_0\,$ is nothing
but typical distance between neighboring gas particles.

For preliminary characterization of cumulants $\,W_{k+1}\,$ it is
useful to notice also that both the initial conditions (\ref{icw})
and equalities (\ref{wp}) are manifestations of the mentioned general
property of cumulants:\, since at $\,t=0\,$ position of the test
particle is non-random (certainly equals to $\,\bm{r}\,$), any joint
cumulant it enters must disappear (similarly, all
$\,\langle\,x,y,...\,,y\,\rangle =0\,$ if $\,x\,$ is a constant),
while (\ref{wp}) as well reflects non-randomness of event ``test
particle is somewhere''.

\begin{widetext}

\,\,\,

\section{The Boltzmann-Grad gas}
Next, consider cumulants $\,W_{k+1}\,$ under the Boltzmann-Grad
limit, when their role becomes especially distinct.

At $\,g\rightarrow 0\,$, the estimates (\ref{ups0}), (\ref{fl0}),
(\ref{ups}) and, very importantly, (\ref{hyp}), lead to the
asymptotic
\begin{eqnarray}
\mathcal{F}\{t,\phi\,|\bm{r}\}\,\rightarrow\,\exp\left [\,\nu_{\,0}
\int\phi(\bm{r})\,d\bm{r} \right ]\,\,\,\,,\,\,\,\,\,\,\,\,\frac
{\mathcal{F}\{t,\phi\,|\bm{r}\}}{\mathcal{F}\{\phi\}}\,\rightarrow
\,1\,\,\label{lim}
\end{eqnarray}
As the consequence, firstly, identity (\ref{id}) simplifies to
\begin{eqnarray}
\int\mathcal{W}\{t,\bm{r}^{\,\prime}|\phi\,|\bm{r}\}
\,\,d\bm{r}\,=\,1\,\,\,\,,\,\,\,\,\,\,\,\,\text{that is}
\,\,\,\,\,\,\,\,\,\int\,
W_{k+1}(t,\bm{r}^{\,\prime}|\bm{r}_1,...\,,\bm{r}_k|\bm{r})
\,\,d\bm{r}\,=\,0\,\,\,\label{id1}
\end{eqnarray}
Secondly, our main formula (\ref{rfunc}), or (\ref{rfunc1}), turns
into
\begin{eqnarray}
\frac {\nu(t,\bm{r})}{\nu_{\,0}}-1\, =\,\int
\phi(\bm{r}^{\,\prime})\,
\mathcal{W}\{t,\bm{r}^{\,\prime}|\phi\,|\bm{r}\}
\,\,d\bm{r}^{\prime}\,\,\,\label{rfuncl}
\end{eqnarray}
Thirdly, definition (\ref{wdef}) of the cumulants
$\,W_{k+1}(t,\bm{r}^{\,\prime}|\bm{r}_1,...\,,\bm{r}_k|\bm{r})\,$
degenerates to
\begin{eqnarray}
W_1(t,\bm{r}^{\,\prime}-\bm{r})+ \sum_{k\,=1}^{\infty}\frac
{1}{k!}\int
F_{k+1}(t,\bm{r}^{\,\prime}|\bm{r}_1,...\,,\bm{r}_k|\bm{r})
\prod_{j\,=1}^k\varphi(\bm{r}_j)\,d\bm{r}_j\,=
\,\,\,\,\,\,\,\,\,\,\,\,\,\,\,
\,\,\,\,\,\,\,\,\,\,\,\,\,\,\,\,\,\,\,\,\,\,\nonumber \\
=\, \left\{\,W_1(t,\bm{r}^{\,\prime}-\bm{r})+
\sum_{k\,=1}^{\infty}\frac {1}{k!}\int
W_{k+1}(t,\bm{r}^{\,\prime}|\bm{r}_1,...\,,\bm{r}_k|\bm{r})
\prod_{j\,=1}^k\varphi(\bm{r}_j)\,d\bm{r}_j\,\right\}\,\exp\left
[\,\int\varphi(\bm{r})\,d\bm{r} \right ]\,\,\,,\label{was}
\end{eqnarray}
where we introduced new dummy function (series expansion variable)\,
$\,\varphi(\bm{r})\equiv\, \nu_{\,0}\,\phi(\bm{r})\,$.

Fourthly, functional
$\,\mathcal{W}\{t,\bm{r}^{\,\prime}|\phi\,|\bm{r}\}\,$ should be
reconsidered in the course of the Boltzmann-Grad limit,
\begin{eqnarray}
\mathcal{W}\{t,\bm{r}^{\,\prime}|\phi\,|\bm{r}\}\,=\,
W_1(t,\bm{r}^{\,\prime}-\bm{r})\,+\,\lim_{\nu_{0}\,\rightarrow
\,\infty }\,\,\,\sum_{k\,=1}^{\infty}\frac {\nu_{\,0}^{\,k}}{k!}\int
W_{k+1}(t,\bm{r}^{\,\prime}|\bm{r}_1,...\,,\bm{r}_k|\bm{r})
\prod_{j\,=1}^k\phi(\bm{r}_j)\,d\bm{r}_j\,\,\,\,,\label{wlim}
\end{eqnarray}
\end{widetext}
as a result of competition between infinitely growing factor
$\,\,\nu_{\,0}\,$,\, on left hand, and infinitely shrinking radius
\,\,$\,l_0\,$\, of decay of
\,$\,\,W_{k+1}(t,\bm{r}^{\,\prime}|\bm{r}_1,...\,,\bm{r}_k|\bm{r})\,$'s\,
via \,$\,|\bm{r}_j-\bm{r}^{\,\prime}\,|\,$\, dependence, on right
hand.

Among three qualitatively different variants of the competition, i.e.
$\,\nu_{\,0}l^{\,3}_0\rightarrow 0\,$\,\,\, or
$\,\nu_{\,0}l^{\,3}_0\rightarrow \,$const\,\, or
$\,\nu_{\,0}l^{\,3}_0\rightarrow \infty\,$,\,\, in fact the middle
variant is acceptable only, since, from the point of view of the
density relaxation, $\,\,\nu(t,\bm{r})/\nu_{\,0}\,\,$\,, the third
variant would mean that none initial density perturbation can be
considered as a weak one (even if it is infinitesimally small) while
the first variant would lead to purely linear relaxation of the
perturbation (even if it is arbitrary large). Hence, once again we
come to conclusion that $\,l_0\,\sim\,1/\nu_{\,0}^{1/3}\,$ (and
without loss of generality we can put on
$\,l_0\,=\,1/\nu_{\,0}^{1/3}\,$).

In order to properly perform the limits in (\ref{wlim}), one should
take in the mind identities (\ref{wp}) (and then (\ref{id1}) as well)
which are very important since say that results of integrations in
(\ref{wlim}) must represent full derivatives with respect to
$\,\bm{r}^{\,\prime}\,$ (and finally also with respect to
$\,\bm{r}\,$).

Then, under assumption that $\,\phi(\bm{r})\,$ is not ``too
arbitrary'' but sufficiently smooth field, quite natural expectation
for the limit of $\,(k+1)$-th term of (\ref{wlim}) reads
\begin{widetext}
\begin{eqnarray}
\lim_{\nu_{0}\,\rightarrow \,\,\,\infty }\,\nu_{\,0}^{\,k}\int
W_{k+1}(t,\bm{r}^{\,\prime}|\bm{r}_1,...\,,\bm{r}_k|\bm{r})
\prod_{j\,=1}^k\phi(\bm{r}_j)\,d\bm{r}_j\,=\,-\,\frac
{\partial}{\partial \,\bm{r}^{\,\prime}}\cdot
\bm{I}_{k+1}(t,\bm{r}^{\,\prime}-\bm{r})\,\,
\phi^{\,k}(\bm{r}^{\,\prime})\, \,\,\,\,\,\,\label{limap}
\end{eqnarray}
with vector functions $\,\bm{I}_{\,k+1}(t,\bm{R})\,$ formally just
defined by this expression. Substituting it into (\ref{wlim}) and
then (\ref{wlim}) to (\ref{rfuncl}) and twice applying integration by
parts, we obtain
\begin{eqnarray}
\frac {\nu(t,\bm{r})}{\nu_{\,0}}-1\,\, \rightarrow\,\, \int
W_1(t,\bm{r}^{\,\prime}-\bm{r})\,\phi(\bm{r}^{\,\prime})\,d\bm{r}^{\,\prime}\,+
\,\sum_{k\,=1}^{\infty}\frac {1}{(k+1)!}\int
w_{k+1}(t,\bm{r}^{\,\prime}-\bm{r})\,\,
\phi^{\,k+1}(\bm{r}^{\,\prime})\,d\bm{r}^{\,\prime}
\,\,\label{rfuncll}
\end{eqnarray}
\begin{eqnarray}
\text{with one more set of (scalar) functions defined
by}\,\,\,\,\,\,\,\,\,\,\,\,\,\,\,\,\,\,\,\,\,
\,\,\,\,\,\,\,\,w_{\,k+1}(t,\bm{R})\,\equiv\,-\,\partial
\bm{I}_{k+1}(t,\bm{R})/\partial \bm{R}\,\,\,\,\,\,\,\,\,\,\,\,\,\,
\,\,\,\,\,\,\,\,\,\,\,\,\,\,\,\,\,\,\,\,
\,\,\,\,\,\,\,\,\,\,\,\,\,\,\,\,\,\label{wf}
\end{eqnarray}
\end{widetext}
According to these definitions and again to identities (\ref{wp}) and
(\ref{icw}), with addition of (\ref{id1}), the new functions must
obey restrictions
\begin{eqnarray}
\int w_{\,k+1}(t,\bm{R})\,\,d\bm{R}\,=\,0\,\,\,\,,
\,\,\,\,\,\,\,\,\,w_{\,k+1}(0,\bm{R})\,=\,0\,\,\,\,,
\,\,\,\,\,\,\,\,\label{wpl}\\
\int \bm{I}_{\,k+1}(t,\bm{R})\,\,d\bm{R}\,=\,0\,\,\,\,,
\,\,\,\,\,\,\,\,\,\bm{I}_{\,k+1}(0,\bm{R})\,=\,0\,
\,\,\,\,\,\,\,\,\,\label{ipl}
\end{eqnarray}
We passed over in silence that in fact transition from left-hand to
right-hand side of (\ref{limap}) needs in one more property of
cumulants
$\,\,W_{k+1}(t,\bm{r}^{\,\prime}|\bm{r}_1,...\,,\bm{r}_k|\bm{r})\,$,
namely, that they remain bounded under the Boltzmann-Grad limit. In
terms of corresponding distribution functions,
$\,F_{k+1}(t,\bm{r}^{\,\prime}|\bm{r}_1,...\,,\bm{r}_k|\bm{r})<\,$const\,.
Orally, magnitude of dynamical inter-particle correlations does not
increase when their mean radius $\,l_0\,$ decreases down to zero.
From the point of view of the BBGKY equations, hardly this statement
gives rise to doubts. What is for a scale of smoothness of
$\,\phi(\bm{r})\,$ needed in (\ref{limap}) when $\,l_0\rightarrow
0\,$ , possibly, it must stay at the level of mean free path
$\,\lambda\,$.

\section{Discussion and resume}
It seems quite obvious that just now considered distribution
functions
$\,\,F_{k+1}(t,\bm{r}^{\,\prime}|\bm{r}_1,...\,,\bm{r}_k|\bm{r})\,$,
corresponding cumulant functions
$\,\,W_{k+1}(t,\bm{r}^{\,\prime}|\bm{r}_1,...\,,\bm{r}_k|\bm{r})\,$
and their children $\,w_{\,k+1}(t,\bm{R})\,$ are in very close
conceptual and mathematical connections with the special full phase
space distribution functions
$\,A_n(t,\bm{R},\bm{v}_1,...\,,\bm{v}_n)\,$, considered in \cite{i1}
in the framework of the ``collisional approximation'' to BBGKY
equations (in \cite{i2} they were renamed to
$\,F_n(t,\bm{R},\bm{v}_1,...\,,\bm{v}_n)\,$), and their children like
$\,W_n(t,\bm{R})\,$ introduced and investigated in \cite{p1}. At
that, equalities (\ref{wpl}) and (\ref{ipl}) show that, in opposite
to $\,W_1(t,\bm{R})\,$, functions $\,w_{\,k+1}(t,\bm{R})\,$ can not
be treated as spatial probability distributions. The lowercase letter
just accentuates this difference between $\,w_{\,k+1}(t,\bm{R})\,$
and true (non-negative) distribution functions $\,W_n(t,\bm{R})\,$.
By implication, however, $\,w_{\,k+1}(t,\bm{R})\,$ and
$\,W_{\,k+1}(t,\bm{R})\,$ reflect similar properties of statistical
ensemble. More careful analysis, with the help of the BBGKY
equations, shows that sooner the first represent time derivative of
the second.

Detailing of these connections is interesting but nontrivial subject
for separate work. At present, the thing what is most important for
us is that in essence both the old and new of the enumerated
mathematical objects represent the same dynamical multi-particle
correlations. This notion, as combined with formally exact relation
(\ref{rfunc1}) (or its limit form (\ref{rfuncll})), leads to
principal conclusion that the specific multi-particle distribution
functions $\,A_n(t,\bm{R},\bm{v}_1,...\,,\bm{v}_n)\,$,
half-intuitively introduced and argued in \cite{i1} as autonomous
complementary (in addition to the one-particle distribution)
characteristics of spatially non-uniform statistical ensemble, now
appear as formally well-grounded concepts of statistical kinetics.

The salt of such statistical characteristics of gas as
$\,A_n(t,\bm{R},\bm{v}_1,...\,,\bm{v}_n)\,$ or
$\,w_{\,k+1}(t,\bm{R})\,$ is that they were irretrievably lost by
conventional kinetics based on the Boltzmann's ``Stosshalansatz'' or,
in other words, ``molecular chaos'' hypothesis. Due to it, in the
Boltzmannian kinetics (at least under the Boltzmann-Grad limit) the
one-particle distribution function ascended the throne as the only
and exhaustive characteristics of gas state, since none
inter-particle correlations have survived. But now, at last, we have
visual and, in my opinion, rather convincing demonstration of
falseness of such kind of kinetic theories.

Concretely, to resume, we demonstrated existence and significance of
dynamical multi-particle correlations what follow from the above
derived chain of relations (\ref{rfunc})$\,\rightarrow \,$
(\ref{rfunc1})$\,\rightarrow
\,$(\ref{rfuncl})+(\ref{wlim})$\,\rightarrow \,$(\ref{rfuncll}).
These relations connect, on one (left) hand,

coefficient functions in series expansion of space-time-varying
density of spatially non-uniform and thus thermodynamically
non-equilibrium gas over its initial density perturbation,\, and, on
the other (right) hand,

cumulant correlation functions which describe creation and evolution
of joint many-particle correlations in thermodynamically equilibrium
gas which is spatially uniform in thermodynamical sense but at the
same time non-uniform in statistical sense owing to ``informational
perturbation'' of gas by information about initial position of one of
its particles (termed ``test particle'').

On the left, initial density perturbation can be arbitrary large,
therefore generally it causes strongly nonlinear relaxation process,
may be even producing shock waves. This means that on the right-hand
side all many-particle correlations really exist (remaining
essentially different from zero even under Boltzmann-Grad limit).

At it was shown earlier, first in \cite{i1} (see also \cite{i2}) and
then in \cite{p1,p2}, these correlations are not things in themselves
but in fact govern a true evolution of the one-particle distribution
function (referred to the ``test particle'') and thus determine
statistics of self-diffusion (Brownian motion) of gas particles, at
that involving possibility of what can be characterized as
``violation of the law of large numbers'' and ``1/f\, fluctuations in
particles diffusivity and mobility''. Immediate cause of the loss of
these correlations at very beginning of kinetics was the way of
thinking when one identifies actual independency of particles or
events, in the sense of dynamics, and their formal independency, in
the sense of mathematical probability theory (see Introduction). The
``molecular chaos'' hypothesis is natural product of such thinking
about ``collisions'' of particles. However, this hypothesis
immediately becomes inadmissible if ``collisions'' are presumed to
satisfy such principal property of dynamics as conservation of phase
volume and consequently conservation of probabilities (for more
explanations see \cite{i1,i2} and references therein). The pay for so
accurate definition of collisions is absence of definite limits for
their relative frequencies, i.e. absence of a priory ``probability of
(one or another sort of) collisions'' at all (paraphrasing
Bernoulli's words \cite{jb}, probability never turns to certainty).
That is why ``molecular chaos'' hard stays unproved (see article ``On
a derivation of the Boltzmann equation'' by O.\,E.\,Lanford in book
\cite{lm} and related comments in \cite{i2}), and instead of the
single Boltzmann equation we have to deal with infinite chain of
equations for specific distribution functions which together
represent time-scaleless fluctuations in ``probability of
collisions''.

Up to now, such revision of kinetics was concerning statistics of
self-diffusion in equilibrium gas only \cite{i1,i2,p1}. But if
continued to non-equilibrium situations it must somehow touch also
kinetics in its usual sense of the fluid mechanics. Intuition would
like to suggest that probable changes in this field must be not
revolutionary. If that is the case, then it is interesting, for
instance, to clarify what information about the Brownian motion
follows from the mentioned relations if one substitutes there results
of conventional kinetics, e.g. those coming from the Boltzmann
equation (though, to the best of my knowledge \cite{lm}, none exact
spatially non-uniform solution to this equation is known). In any
case, we have found new incentives to critical review of kinetics.




\end{document}